\documentstyle[12pt,epsf,titlepage,fleqn]{article}

\newcommand{\be }{\begin{equation}}
\newcommand{\ee}{\end{equation}}
\newcommand{\ba}{\begin{eqnarray}}
\newcommand{\ea}{\end{eqnarray}}

\newcommand{\non}{\nonumber}

\begin{document}

\begin{titlepage}
\vspace*{0.1truecm}
\hfill{ R/96/28}

\hfill{ hep-th 9610235}

\begin{center}

{\LARGE \bf Properties of Supermembrane Axions}

\vspace{.7cm}

{\large Fermin ALDABE}

 faldabe@phys.ualberta.ca

\vspace{.7cm}

{\large\em DAMTP, University of Cambridge,\\Silver Street,
 Cambridge CB3 9EW,U.K.\\and\\
Theoretical Physics Institute, 
University of Alberta,\\ Edmonton Alberta, T6G 2J1, CANADA}

\vskip 10pt
\begin{quotation}
\baselineskip=1.0em

\vspace{1cm}

{\normalsize

{\bf Abstract}. In the presence of membranes, M-theory becomes in the
low energy limit 11 dimensional supergravity  action
coupled to a supermembrane action.  The fields of the
first action are the same fields which couple to the membrane.
It is shown that the axionic 
moduli of the membrane
obtained by wrapping the three form potential about three-cycles
of a Calabi-Yau manifold
can take nonzero integer values.
This novel property allows M-theory to have
smooth transition from the Kahler cone of a geometrical
phase to a Kahler cone of another geometrical phase.
Nongeometrical phases which define the boundary of the 
extended
Kahler cone of the geometrical phases 
have discrete spectrum,
and are continuously connected to the geometric phases. 
Using this new property, we relate the M-theory model dependent 
axion to the type IIA model dependent
axion and show that a potential develops for the type IIA axion
in the strong coupling regime which does not seem to be generated
by instantons.
Evidence is presented, using these moduli,
which supports the Strominger conjecture on the
winding p-branes.  }

\end{quotation}
\end{center}

\end{titlepage}

\section{Introduction }

It is by now accepted that M-theory has  supermembranes in its spectrum,
and that its has as its low energy limit 11 dimensional
supergravity.  The low energy limit of M-theory in the presence of membranes
is then 11 dimensional supergavity coupled to a supermembrane action.
Moreover, the coupling of the supermembrane to this supergravity action
implies that the fields which appear in the supermembrane action are the
same as those in the supergravity action.

Three issues of M-theory will be dealt with in this paper under
this assumption.
The first
one regards the phases of M-theory in the absence of membranes
\cite{W1} analyzed by Witten.  
The second one regards the generation of a non perturbative potential
for the type IIA axion in the strong coupling limit.
The third one regards the Strominger conjecture on winding p-branes \cite{S}.

\vspace{.5cm}

\noindent
{\em Phases}

\vspace{.5cm}

M-theory on a Calabi-Yau manifold \cite{Fer} can have different phases
\cite{W1}. 
In order to analyze them, it is necessary to define
the Kahler cone of a given Calabi-Yau manifold $X$ as the space of all 
Kahler
metrics of $X$. This cone has a boundary where the size of 
a rational curve shrinks to zero size.   It is also possible to
move to the exterior of the Kahler cone:  a curve defined on the manifold
$X$ will have negative volume in this region, but it is possible to flop the
rational curve so that its volume becomes positive when interpreted as a
rational curve of a manifold $Y$ which is birrationally equivalent to $X$.
Then, by varying the moduli of the Kahler cone, 
we can transition to the set of manifolds which are birrationally 
equivalent to the manifold $X$, and it is thus convenient to
define the extended
Kahler cone, as the union of all Kahler cones of the manifolds which are
birrationally equivalent to
the manifold $X$ and which lead to geometrical phases.  

In \cite{W1}, the behavior of 11 dimensional supergravity
on  the boundary of a Kahler cone not lying on the boundary of the extended
Kahler cone was analyzed.  It was shown that the 5-dimensional Chern-Simons
term signals when a transition from $X$ to $Y$ takes place.  At the 
classical level, this phase transition is sharp since the spectrum
becomes continuous.  Also, it was 
argued that, as opposed
to string theory where one finds abstract phases outside the extended
Kahler cone (such as Landau-Ginzburg
orbifold), in 11 dimensional supergravity the moduli space seems to 
end abruptly, and abstract phases
are absent.  In order to reach these conclusions, it is necessary to make a 
detailed analysis of the Kahler moduli space.  We will see in the next
sections by coupling the supermembrane to 11 dimensional supergravity
and including the moduli space of three-cycles in the
analysis of the phases actually smoothes out all transitions of M-theory
including  those which are 
non geometric phases.

\vspace{.5cm}

\noindent
{\em Axionic Potential}

\vspace{.5cm}

The moduli of these three-cycles in M-theory
are periodic but as we shall show take integer values.
This is to be contrasted with the moduli of string theory obtained by
wrapping the B-field about two-cycles.  The latter moduli are periodic
but do not take integral values.   After compactification of M-theory
on $S^1\times X\times R^4$, it is possible to relate the moduli of three-cycles
in M-theory 
to the moduli of two cycles in type IIA string theory on $X\times R^4$.
The latter are model dependent 
axions of the type IIA string.  In the strong coupling limit,
the type IIA axions become M-theory axions.  This means that as the coupling
constant increases, the  type IIA axions are restricted to certain discrete
values
which implies the existence of a potential.  This potential does not seem
to be generated by instanton effects,  and at present, it is not
possible to estimate the order of magnitude of the mass acquired by
the axions. 

\vspace{.5cm}

\noindent
{\em The Winding Conjecture}

\vspace{.5cm}

In order to interpret the divergent coupling of string theory
near a conifold
singularity,
Strominger \cite{S} postulated that the winding of
p-branes with $p>1$ should
be treated  in a different manner then the winding of strings.
We must require that a p-branes with $p>1$, winding $N>1$ times
be treated as an $N$ particle state.  Otherwise, a p-brane which
wraps $N$ times about a p-cycle of a compactification would contribute to
the beta function in a similar way as a p-brane wrapped once about the 
same cycle.  This, of course,
 ruins the relation between the conifold singularity and 
the renormalization of the coupling.  On the other hand, 
as explained in \cite{BBS},
a string winding $N>1$ times about a cycle must be treated as a single
particle state rather than a multiple particle state because there
are bound states at threshold \cite{DH}.  

In the context of string
theory and without any knowledge of M-theory, discriminating between
the winding properties of type IIA membranes and the type IIA strings
does not pose a problem.  But as soon as M-theory comes into the picture,
some inconsistencies appear.
For example, a membrane wrapped once 
on $S^1$ yields after dimensional reduction a type IIA string \cite{D}.
But what kind of string will a membrane wrapped $n$ times about
an $S^1$ yields?  Given Strominger's conjecture, we expect that the
string obtained does not depend on the number of times the membrane
is wrapped about the $S^1$.  This is equivalent to the requirement
that the string coupling and the axionic charge should not depend on the 
winding number of the membrane about $S^1$.  However, consider a membrane which is wrapped
$n_1$ times about the first  one-cycle of $T^2$ and $n_2$ times about the 
second
one-cycle of $T^2$.  If we use the rule that dimensional reduction 
of a membrane yields the same string, then dimensional reduction of
the membrane about the first one cycle
yields the statement that $N$ windings of the string about the
second cycle
should lead to an $N$ particles state.
This is then in contradiction with the departing conjecture that 
$N$ windings of a string about a one-cycle should yield a single
particle state.  Equivalently, it is 
in contradiction
with the findings of \cite{DH}: an n-wound string should not yield
the same state as $n$ single-wound string.

We will later see that the coupling (M-theory 
axion moduli) of the membrane to the three form
potential $C$ has a different structure than the coupling (type IIA
axion moduli) of the string
to the two form potential $B$.  This difference between M-theory
and type IIA axionic moduli leads to a resolution
of the Strominger winding conjecture.

\vspace{.5cm}

\noindent
{\em Summary and Outline}

\vspace{.5cm}

Although we do not have a good grip on the quantum field theory
describing the fundamental membrane, recent results point out
that the supermembrane action  will be quantized
\cite{R}.  Also, supermembranes seem to be reasonable candidates
whose low energy limit yields 11-dimensional supergravity \cite{T}.

Here we assume that M-theory in its low energy limit is 
11 dimensional supergravity coupled to membranes.  We 
then analyze the phases of
supermembranes propagating on a Calabi-Yau threefold to draw
conclusions on the phases of M-theory.
We will restrict to Calabi-Yau three-folds, although higher
dimensional compactification have also lead to new and interesting
physics in lower dimensions \cite{AK,W5}.
Supermembranes propagating on 
a Calabi-Yau threefold compactification have two types of moduli.  The first
type are those obtained by integrating the Kahler metric over a two-cycle
of the compactification.  These will not play a very important role
in our discussion and most of their aspects have been dealt with in 
\cite{Fer,W1}.  The second type is obtained by wrapping the three-form
potential about a three-cycle of the internal space.  These moduli  will be
our {\em vedette}.  By analyzing the gauged linear sigma model with N=2 supersymmetry
in three dimensions, we will be able to conclude that these moduli 
take discrete values.  These property can be used to show that
in the presence of membranes,
the transition of M-theory
from one manifold $X$ to another manifold $Y$ takes place in a smooth manner.
We will also study the phase transition at the boundary of the
extended Kahler cone and show that non geometrical phases such as Landau
Ginzburg orbifolds, having a discrete spectrum, are present at the boundary
of the extended Kahler cone, and are continuously connected to this cone
when membranes are present.
In addition we will use this novel feature of supermembranes to show the
existence of an axionic  potential in the strong coupling limit and
to provide evidence in favor of  the Strominger conjecture.

The paper is organized as follows.
In section 2 we review the string gauged linear sigma model to explain which 
are the quantum effects which  allow for smooth phase transitions between
strings propagating on birrationally equivalent manifolds.  In Section 3 we
construct the membrane gauged linear sigma model coupled to
11 dimensional supergravity and 
show that quantum effects, absent in the low energy effective theory, also
allow for the smooth phase transitions of M-theory propagating on 
birrationally equivalent 
manifolds.  Section 4 deals with the analysis of phase transitions
between geometrical and nongeometrical phases of M-theory theory.
In Section  5 we show the existence of an 
axionic potential in the strong coupling limit of string theory.
The last section is devoted to providing evidence for
 the Strominger winding conjecture.

\section{Phases of N=2 Strings}

We begin by reviewing the string gauged linear sigma model \cite{W2}
used to describe the 
phase  transitions 
of string theories with N=2 supersymmetry between two birrationally equivalent
manifolds.  The two dimensional
Lagrangian is 
obtained from the dimensional reduction to D=2 N=2 of a D=4 N=1 gauge action
coupled to matter \cite{W2}.  The bosonic sector is\footnote{We follow the
conventions of \cite{W2}}
\ba 
L&=&L_{kin}+L_{gauge}+L_{D}\non\\
L_{kin}&=& \int d^2y\ (\partial_m \bar a_i \partial^m a_i
+ \partial_m \bar b_i \partial^m b_i)\non\\
L_{gauge}&=&\int d^2y\ ( v_{01}^2+\partial_m\sigma\partial^m \bar{\sigma})\non\\
L_D&=&-r\int d^2y D.\label{L2}
\ea
$L_{kin}$ is the kinetic term for four chiral superfields 
$A_i, \ B_i;\ i=1,2$, whose bosonic scalars are $a_i$ and $b_i$ respectively.
$L_{gauge}$ is the kinetic term for the gauge field $v_m\ m=0,1$ ( giving 
rise to the field strength $v_{01}$ ) and $\sigma=\sigma_1+i\sigma_2$,
where $\sigma_i$ are the scalars of the dimensionally reduced D=4 vector
multiplet.  Thus, $v_{m}$ and $\sigma_i$ make up part of 
a D=2, N=2 vector multiplet.
$L_D$ is the 
Fayet Illiopoulos (FI) term and $D$ is the auxiliary field of the D=2 vector
multiplet.   
In addition, we may add a topological term 
\be 
L_{\theta}={\theta}\int d^2y \ v_{01}
\label{topo}
\ee
which will be crucial in our discussion.  The parameter $\theta$ is just the
two dimensional theta-angle which is periodic \cite{RW}
\be
\theta\sim\theta+1.
\ee
It is the expectation value of the 4-dimensional model dependent axion field.

We will consider the example of \cite{W2} where the $A_i$'s  will carry 
positive unit charge with respect to the gauge field
and the $B_i$'s will carry negative unit charge.
This complex fields can be though as coordinates on $C^4$.
 The bosonic potential of this model is obtained by integrating out the
auxiliary field $D$
\be 
U=\frac{e^2}{2}(\sum_i|a_i|^2-|b_i|^2-r)^2+|\sigma|^2(\sum_i|a_i|^2+|b_i|^2).
\ee
For $r>>0$, the $a_i$'s cannot both be zero since the first 
term in the bosonic potential is
proportional to $D^2$.  The values of $a_i$ determine a point in a copy of
$CP_a^1$.  
For $b_i=0$, the gauge symmetry can be used 
to divide $\sum_i|a_i|^2=r$ by $U(1)$ and therefore, $r$ is the Kahler form
of $CP^1_a$.
The zero section of the symplectic quotient, $Z_+$, $Z_+\to CP^1_a$,
is a
genus zero holomorphic curve which is obtained by setting $b_i=0$ \cite{W2}.

For $r\ll0$, the role of the $a_i$'s and the $b_i$'s are exchange, and
the $b_i$'s define a point in $CP_b^1$ with Kahler form $-r$.
The zero section of the  symplectic quotient $Z_-$ is also a genus
zero holomorphic curve.  

As we transition from positive to negative $r$ a change of topology, a flop,
 takes
place because the size of $CP_a^1$ shrinks to zero size and is replaced
by $CP^1_b$.  This means that $Z_+$ is replaced by $Z_-$ and therefore,
we move from a manifold $X$ where $CP^1_a$ has positive volume to a 
birrationally equivalent manifold $Y$ where $CP_b^1$ has positive volume.
In order to arrive to the point $r<<0$, 
we must first transverse the point $r=0$.  At this point in the moduli
space, there are wave functions supported near $b_i=a_i=0$ and $|\sigma|\gg0$
which have a continuous spectrum.  However, when $\theta$ in
the topological term (\ref{topo}) is different from zero, the spectrum remains
discrete, thus insuring a smooth phase transition \cite{W2}.

The term (\ref{topo}) can be combined with the FI term 
\be 
i t\int d^2y (D+iv_{01})
\ee
where $t={\theta}+i r$ is the moduli of the complexified 
metric $J=B+i G$
where $B$ is the antisymmetric tensor and $G$ is the Kahler metric.
Thus, $\theta$ is the coupling of the antisymmetric tensor.  This also follows
from the fact that for $r\gg0$ we may rewrite (\ref{topo}) in the form
\be 
{\theta}\int d^2y \epsilon^{ij}\partial_iX^M\partial_jX^NB_{MN}\label{t2}
\ee
where $B_{MN}=\partial_{[N}A_{M]}$ is the antisymmetric tensor 
which locally is pure gauge, and the fields $X^M$ parametrize a 10 dimensional
space.

\section{Phase Transitions Between Geometric Phases of M-theory}

Is this transition between birrationally equivalent manifold 
also smooth for membranes?  
If so, this implies that in the precence of supermembranes the transitions
of M-theory are smooth.

The parameters 
$r$ and $\theta$ in string theory are scalars of a D=4 N=2 spacetime
vector multiplet.  In the case of 11 dimensional supergravity on a manifold $X$,
the N=2 vector multiplet is five dimensional, and therefore has one scalar
only.  This effectively sets $\theta$ to zero, thereby suggesting only the
presence of sharp phase transitions.   
However, as we shall 
see,  when supermembranes are added to the 11 dimensional supergravity action,
there are  moduli parametrizing the coupling of the
three-form antisymmetric tensor $C_{MNP}$.  
These moduli allow
the phase transitions of M-theory between birrationally
equivalent manifolds to be smooth, just like in string theory.

Consider the dimensional reduction to D=3 N=2 
of the N=1 D=4 gauge action coupled to 
matter.  The dimensional reduction of this theory down to D=2
will yield the string gauged linear sigma model.  This will guarantee
that the model describes supermembranes whose dimensional reduction on $S^1$
yields strings.  The D=3 N=2 action will
yield the membrane gauged linear sigma model.
The bosonic sector of the Lagrangian is 
\ba 
L&=&L_{kin}+L_{gauge}+L_{D}\non\\
L_{kin}&=& \int d^3y\ (\partial_m \bar a_i \partial^m a_i
+ \partial_m \bar b_i \partial^m b_i)\non\\
L_{gauge}&=&\int d^3y\ ( v_{mn}^2+\partial_mS\partial^m S)\non\\
L_D&=&-r\int d^3y D.\label{L3}
\ea
The action of the chiral fields which depend on a three dimensional
space is the same as those of the two dimensional action (\ref{L2}).
The same occurs for the D-term.  However, the content of the D=3, N=2 
vector multiplet is different from its two-dimensional analog because
there is only on scalar $S$.
The bosonic potential of the
D=3 action is also obtained after integrating out the auxiliary field $D$
\be 
U=\frac{e^2}{2}(\sum_i|a_i|^2-|b_i|^2-r)^2+S^2(\sum_i|a_i|^2+|b_i|^2).
\ee
For $r>>0$ and demanding preservation of N=2 supersymmetry, we obtain
an action for a supermembrane propagating on a manifold $X$, which has a 
symplectic quotient $Z_+$.  For $r\ll0$, we obtain the action for a 
supermembrane propagating on a manifold $Y$ which is birrationally equivalent 
to the manifold $X$ and has a  simplectic quotient $Z_-$.

We are not able to go from positive to
negative $r$ without encountering a sharp phase transition because there is
no theta-angle to prevent the spectrum from becoming continuous for wave
functions supported near $a_i=b_i=0$ and $S\gg0$.
In order to obtain smooth phase transitions, we must couple the
supermembrane to the 
3-form antisymmetric tensor $C$ which is part of the massless
spectrum of D=11 supergravity.  This coupling also insures k-symmetry
of the supermembrane action \cite{T,HP}. This is done by adding  a term to the 
D=3 Lagrangian (\ref{L3}) whose bosonic sector is 
\be 
m\int d^3y (\epsilon^{ijk}A_i\partial_jA_k+S^2).\label{nose}
\ee
This term is particular to three dimensions:  it is a mass term for the
worldvolume U(1) gauge multiplet 
$(A_i,S)$ which preserves gauge symmetry as well as
N=2 supersymmetry.  As opposed to the topological coupling $\theta$ 
in (\ref{topo})
which is periodic,
the coupling $m$ takes values on ${\bf Z}$ \cite{W4}\footnote{I thank
E. Witten for pointing this out to me.} and it is the coupling of the
membrane to the three form potential $C$.

Now we may consider the action for a supermembrane which has a rational
curve on a manifold $X$ which can be flopped to a rational curve of a 
manifold $Y$. 
For $r\gg0$, we encounter the worldvolume 
action on a rational curve on a manifold $X$.  For $r\ll0$ we encounter the
worldvolume action on a rational curve on a manifold $Y$ which is 
birrationally equivalent to the  manifold $X$.   In order to go from $X$ to $Y$
we must pass through the point $r=0$.  At this point, if $m=0$ we find
wave functions supported near $a_i=b_i=0$ and $S\gg0$ which have a 
continuous spectrum and therefore
a singularity at $r=0$.  However, if $m\ne0$ such wave functions are absent
because $S=0$,
and we are guaranteed a smooth phase transition in going from $X$ to $Y$.
The topological term in the two dimensional action is analogous to the
topological term of the three dimensional action.  Since the former 
yields stringy effects which cancel the continuous spectrum, the latter
should also be though as a membrany quantum effect. 
Thus, M-theory has smooth phase transitions like string theory
when transversing a boundary of a Kahler cone provided quantum effects are
included in the effective membrane action.

We see that different values of $m$ lead
to different physical situations.  In particular, a membrane
with $m=0$ does not
yield smooth transitions while a membrane with $m\ne0$ do.
Thus, the two membranes are different and yield different phases for
M-theory.

\section{Phase Transitions Between Geometric and Nongeometric
Phases of Supermembranes}

So far we have considered the phases of M-theory within the boundary
of the extended Kahler cone (the set of Kahler cones which yields
geometric phases) and on a local patch containing an inner boundary
which does not belong to the boundary of the extended Kahler cone.
We would now like to show that the D=3 Lagrangian indeed describes a 
membrane propagating on a Calabi-Yau manifold, and also analyze the transition
of M-theory 
to a Landau-Ginzburg orbifold phase which, as explained in \cite{W1},
is a cone of zero size, and therefore defines the boundary of the extended
Kahler cone. 

For simplicity, we will consider a Calabi-Yau manifold given by 
the quintic polynomial embedded on $CP^4$.
In order to define  a membrane on  this Calabi-Yau manifold, we must consider
the action whose bosonic sector is\footnote{Here we also
use the conventions of \cite{W2}} 
\ba 
L&=&L_{kin}+L_{gauge}+L_{D}+L_{W}\non\\
L_{kin}&=& \int d^3y\ (\partial_m \bar \phi_i \partial^m \phi_i
+ \partial_m \bar p \partial^m p)\non\\
L_{gauge}&=&\int d^3y\ ( v_{mn}^2+\partial_mS\partial^m S)
+m\int d^3y (\epsilon^{ijk}A_i\partial_jA_k+S^2)\non\\
L_D&=&-r\int d^3y \ D\non\\
L_{W}&=&\int d^3y\  F_i \frac{\partial W}{\partial\phi_i}+
F_p \frac{\partial W}{\partial p}.\label{ef3}
\ea
The chiral fields are $\Phi_i\ ;i=1,...,5$, and $P$,  their bosonic
scalars are $\phi_i$ and $p$ respectively , and their auxiliary fields
are $F_i$ and $F_p$ respectively.  The chiral fields $\Phi_i$ and $P$
will have charge
$+1$  and $-1$ respectively with respect to the U(1) field.
The new structure in the Lagrangian is the term $L_W$  which is the
potential of the model.  It has the form
\be
W=p Q(\phi)
\ee
where $Q$ is a homogeneous degree five polynomial.  It is also
transverse, so that its first derivatives with respect to the $\phi$'s
vanish simultaneously only when all $\phi$'s vanish.
Integrating out the
auxiliary fields $F_i$, $F_p$ and $D$, we obtain the bosonic potential
\ba
U&=&\frac{e^2}{2}(\sum_i|\phi_i|^2-|p|^2-r)^2+S^2(\sum_i|\phi_i|^2+|p|^2)
\non\\
&&+|Q(\phi)|^2+|p|^2\sum_i|\frac{\partial Q}{\partial \phi_i}|^2+m S^2.
\ea
For $r\gg0$ we see that the $\phi$'s cannot all vanish.  Then, by the
transversality of the polynomial $Q$, $p$ must vanish.  The vanishing
of the first term in the bosonic potential yields an $S^9$ which after
moding out by the U(1) gauge symmetry becomes a $CP^4$ with Kahler class
proportional to $r$.  
The condition that
$Q$ vanish in order to minimize the potential, defines an
embedding of the quintic polynomial $Q$ on $CP^4$ which yields a
Calabi-Yau manifold: the quintic.
As in string theory, the scalar $S$ and the gauge field both become massive.
Thus the effective theory for large $r$ is a supermembrane propagating
on the quintic.  This is to be compared with the phase analysis done
for strings in \cite{W2}.  They both have the same behavior for large $r$,
and dimensional reduction of the effective action (\ref{ef3}) on $S^1\times X$
yields a string
propagating on  $X$.

The case in which $r\ll0$ is also similar to that of strings.  There, $p$
cannot vanish and transversality of $Q$ implies that all $\phi$'s must 
vanish; the U(1) vector multiplet becomes massive.  The massless theory
is a Landau-Ginzburg orbifold (LGO) for the membrane.  
It is not possible to make a statement of the infrared and ultraviolet
behavior because the action is nonrenormalizable.

\vspace{1cm}

\begin{figure}[hb]

\includegraphics{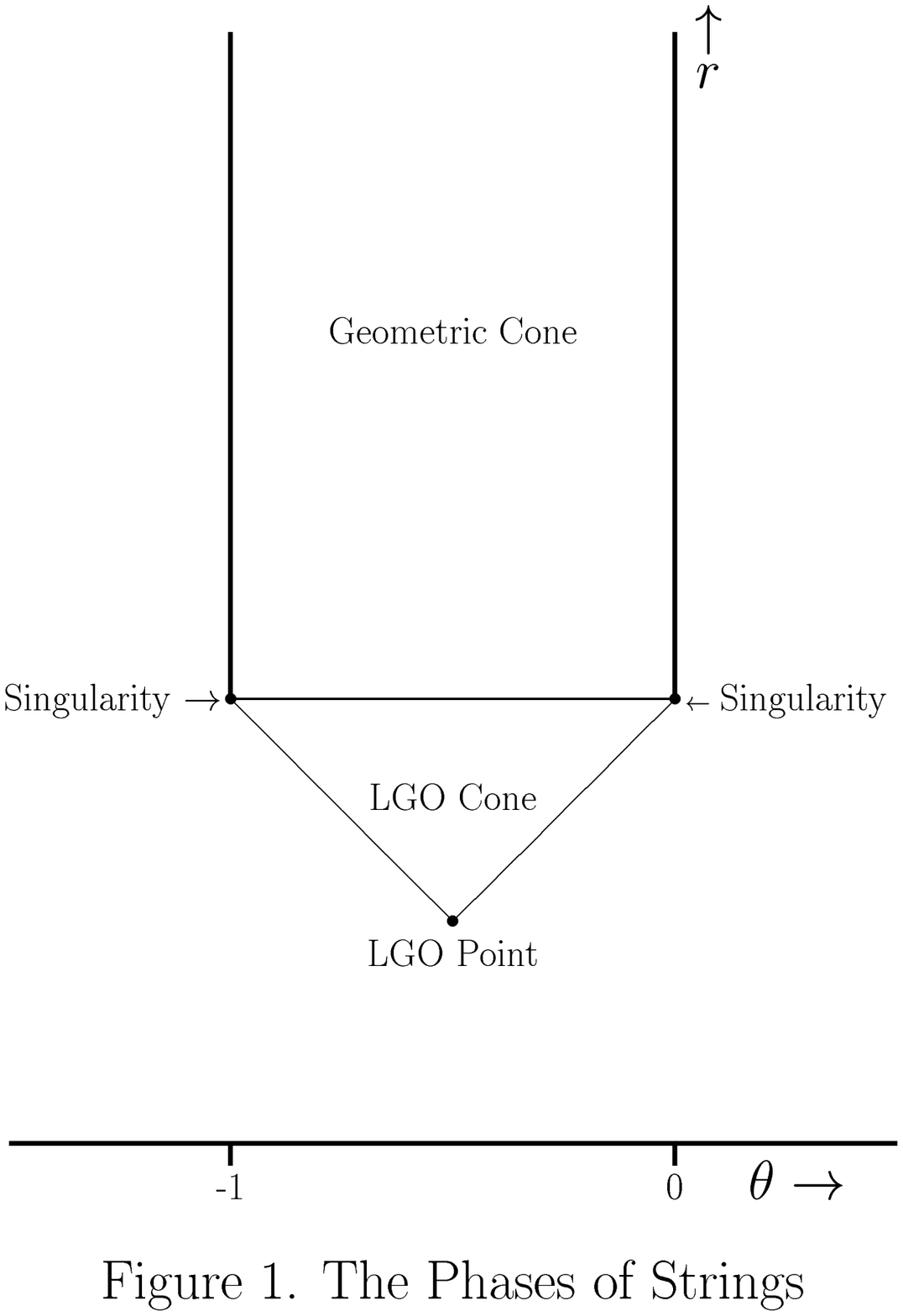}
\includegraphics{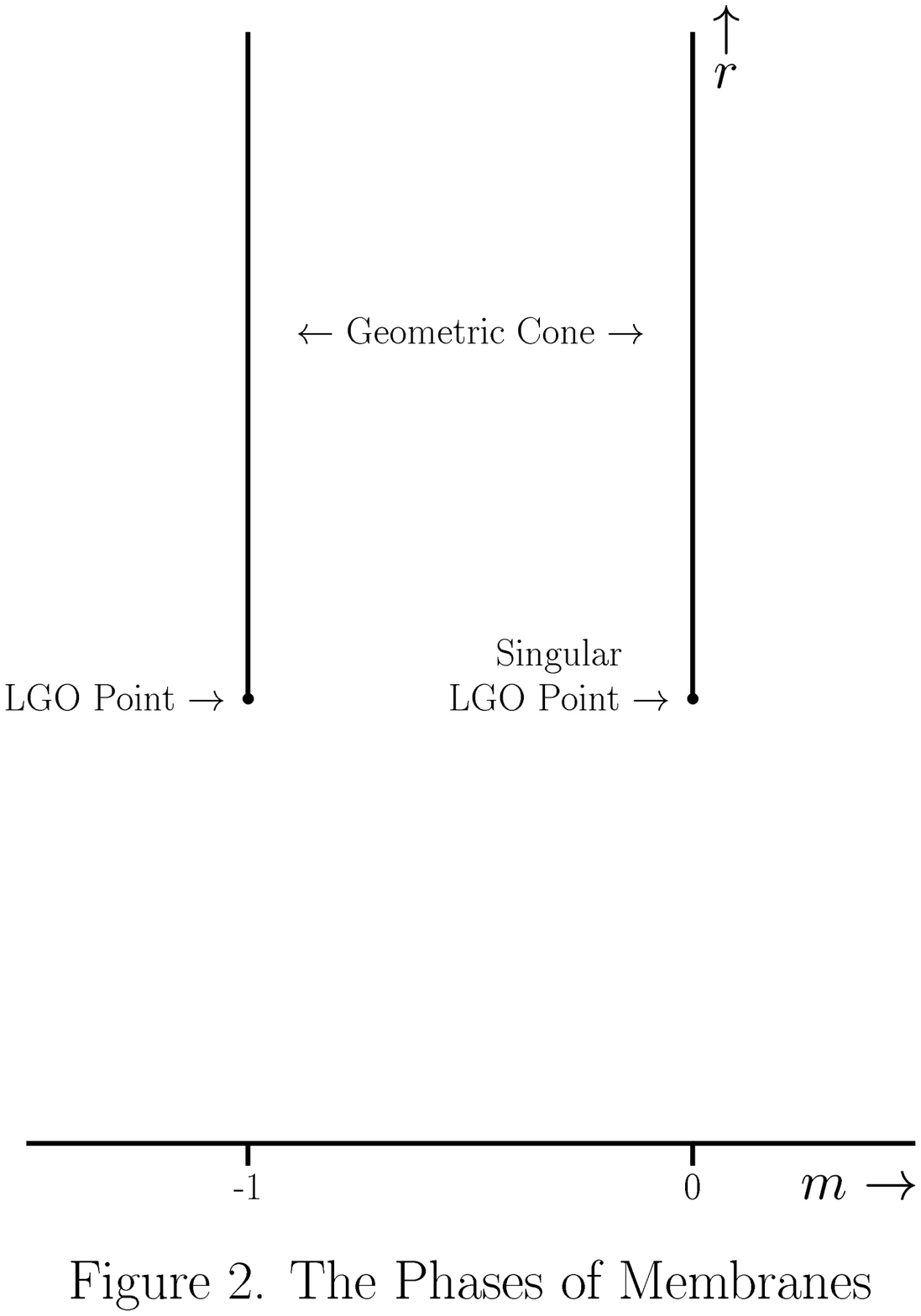}
\vspace{11.0cm}

\label{supergraph}
\end{figure}

So far,
we have been working with linear sigma model coordinates.  
In string theory, for positive $r$ the cone in special coordinates is 
of the same size as in linear sigma model coordinates
because instanton effects are small.  However, for negative
$r$, where instanton effects are large,
the LGO cone in special coordinates is squashed to a thickness of order
$\alpha'$.  This is shown in Figure 1, where the LGO phase 
is the triangle formed
by the LGO point and the two singularities.
The Kahler class of the LGO cone in type IIA string theory can
be related to the Kahler class of the LGO cone in supermembrane theory and
the radius $R$ of compactification of the eleventh dimension \cite{W1}
\be
K_M=\frac{K_{II}}{T^{\frac{1}{3}} R}
\ee
where $T$ is the membrane tension.
This relation implies that in the limit in which $R\to\infty$, 
or equivalently $\alpha'\to0$,
the Kahler cone of the 
 LGO phase of the supermembrane, 
is squash to the boundary of the extended Kahler cone.  The moduli
space of the supermembrane is shown in Figure 2.

For $m\ne0$, the LGO phase forms a boundary of the extended Kahler cone of 
geometric phases. The LGO point which lives on this boundary is 
continuously connected to the extended Kahler cone of geometric phases
and has a discrete spectrum, thus making it a well defined phase in
M-theory.   Notice that once more, different values of $m$ lead
to different physical situations.  In particular, a membrane
with  $m=0$ has a singular LGO point with continuous spectrum
 while a membrane with $m\ne0$ has an LGO phase with discrete spectrum.
Again, these two membranes have different phases and lead to 
different physics. 

\section{Axionic Potential of the Membrane on a 3-Fold}

We have seen in the two previous sections that the moduli obtained
by wrapping the three-form potential $C$ about a three-cycle take integral
values.   We have also seen that the physics for different non zero 
values of these moduli is the same and that the physics for non zero 
and zero values of the moduli is different.  For non zero values 
the membrane has smooth transitions between Kahler cones and also has
smooth transitions to well defined Landau Ginzburg phases.
For zero values of these moduli we find that the Landau Ginzburg phases
is singular and the transition between different Kahler cones is
sharp. 

In this section we will extend the properties of these moduli to those
moduli which are also obtained by wrapping the three form potential $C$
about a three-cycle  which can be written as a product of a two-cycle and
a one-cycle.  This mild assumption, which is based on the fact that there
is no underlying reason why the membrane should distinguish between a simply
connected and a non simply connected cycle, will allow us to perform a
dimensional reduction of 11 dimensional supergravity to 10 dimensional 
type IIA supergravity and relate the moduli of the $C$ field to the
axion of type IIA supergravity obtained by wrapping its two-form potential
$B$ about a two cycle.  This relation between the moduli of the $C$ field,
which we will refer to as M-theory axion $a_M$, and the usual axion
which we will refer to as type IIA  axion $a_S$ will serve to establish
one property: 
the existence of a 
potential for the axion in the strong coupling limit which does not
depend on instanton contributions.

We first show that the axion $a_M$ is periodic by closely following
the analysis of Rhom and Witten \cite{RW} to show the
periodic properties of $a_S$.  Note that
the term 
\be
I=\int_W d\Sigma^{NMPQ}G_{NMPQ}
\ee
present in the supermembrane action does not need to be single-valued.
Here $W$ is  a closed four manifold and 
\be
G=dC.
\ee
Rather, it is $e^{iI}$ which must satisfy this property.  This implies that 
$I=2\pi n$ where $n\in {\bf Z}$

We now define $a_M$ to be
\be
a_M(x^{\mu})=\int_T d\Sigma^{MNP}C_{MNP}
\ee
where the integral is over the closed three manifold.
Then, in circling a loop $\gamma$ at fixed transverse
distance from a membrane which we will parametrize by the angular coordinate
$\phi$, the change in $a_M$ is 
\be
\delta a_M=\int_0^{2\pi} d\phi\frac{da_M}{d\phi}=\int_{\gamma\times T} dC
=\int_{\gamma\times T} G.
\ee
Since ${\gamma\times T}$ is a closed four manifold we find that
\be
\delta a_M=2\pi n
\ee
 and therefore $a_M$ is a periodic variable.

Having established the periodicity of $a_M$ we continue by 
considering 11 dimensional supergravity to show the existence
of an axionic potential.  The bosonic
sector of the action is 
\be
\int d^{11}x( \sqrt{-g}(R+G^2)+G\wedge G\wedge C)\label{q0}
\ee
We will dimensionally 
reduce this action to four dimensions.  The internal space used to
reach four dimensions is $X\times S^1$ where $X$ is a Calabi-Yau threefold.
A scalar field $a$ 
will be an axion only if it satisfies that it is periodic and if
it exhibits a  coupling of the form
\be
\int d^4x \ a \ F\tilde{F}\label{q1}
\ee
where $F$ is a two-form field strength and $\tilde{F}$ its dual. 
A term of the form (\ref{q1}) can only  be obtained from the topological
term in (\ref{q0}).  The two-form field strength is obtained by wrapping
$H$ about a two-cycle found in $X$
\be
H_{\mu\nu mn}=F_{\mu\nu}(x^{\rho})\wedge b^I_{mn}(y^s)
\ee
where $\mu,\nu$ label space time coordinates, $m,n$ label 
internal coordinates of $X$ and $b^I$ is a two-form on $X$.
We have two choices 
to obtain a scalar from the $C$ field.  The first and less interesting is
by wrapping the $C$ field about a three-cycle of $X$, the second and more
interesting for our discussion is to wrap $C$ about a two-cycle of $X$ and
the one-cycle of $S^1$ which will lie in the 11$^{th}$ dimension
\be
C_{mn11}=a_M^I(x^{\mu}) b^I_{mn}(y^s,x^{11})\wedge b_{11}(y^s,x^{11})
\ee
where $a_M$ will be the M-theory
axion and $b_{11}$ is the one-form on $S^1$.  $b^I$
is again a two-form on $X$.

With this ansatz, the $GGC$ term in (\ref{q0}) takes the form
\be
\int d^{4}x\ a_M^K F^J\tilde{F}^I\int d^6y\ dx^{11}\ b_{11}\wedge
 b^I\wedge b^J\wedge
b^K\label{r1}
\ee

Thus the scalar $a^K_M$ is the axion of M-theory.  As we saw in the previous 
sections, $a^K_M$  takes integral values.  This is not the case of the axion of 
type IIA supergravity.  

The type IIA axion is obtained from the topological
term of type IIA supergravity
\be
\int d^{10}x\ G\wedge G\wedge B\label{o1}
\ee
where $G$ is the four-form field strength of type IIA supergravity and
$B$ is the two for potential of type IIA supergravity.
Substituting into (\ref{o1}) the ansatz
\ba
G_{\mu\nu m n}&=& F^I_{\mu\nu}(x^{\mu})\wedge b^I_{mn}(y^s)\non\\
B_{mn}&=& a_S^I(x^{\mu}) b^I_{mn}(y^s)
\ea
where $b^I$ is a two-form on $X$  we obtain
\be
\int d^{4}x\  a_S^K F^J\tilde{F}^I\int d^6y\ b^I\wedge b^J\wedge
b^K\label{r2}
\ee
A close look at (\ref{r2}) reveals that $a_S$
has an axion like coupling.  Furthermore, as we have seen in this section, 
$a_M$ is periodic and so is  
$a_S$ \cite{RW}.
Thus, both $a_M$ and $a_S$ behave like axions which also depend on the
type of compactification and are therefore model dependent axions.
Another property they share is that $a_M^I$ is the strong coupling of
$a_S^I$.  Indeed, in the limit that the $S^1$ parametrized by $x^{11}$ is
very large, we are in the M-theory regime which is the strong coupling limit
of type IIA supergravity.  In this strong coupling limit, $a_M$ is the
axion.  When the $S^1$ is very small, M-theory collapses to type IIA 
supergravity and $a_M$ collapses to $a_S$.  This follows from the
fact that for very small $S^1$ we can take the fields tangent to $X$
to be independent of
$x^{11}$  and then (\ref{r1}) must collapse to (\ref{r2}).  This can
only happen  provided we identify $a_M$ with $a_S$ in this limit.

This identification between $a_M$ and $a_S$ along with what we have
learned in the previous section about $a_M$ imply that a potential
for the axion is generated in the strong coupling limit.
In the strong coupling regime, $a_M$ can only take integral values.
We have further shown that while $a_M \ne0$ and $a_M =0$ lead to different
physical behavior of the membrane, we cannot really distinguish between
the different possible values of $a_M$ which are different from zero.
Thus we can use the identification 
\be
a_M\sim a_M+1
\ee
when $a_M\ne0$.  Therefore, we have to possible values: $a_M=0,1$.

Now consider the situation in which we start in the weak limit of M-theory
(type IIA supergravity).  In this case, the axion is $a_S$ and it can
take any value between $0$ and $1$.  As we increase  the coupling constant
and reach the M-theory regime, we find that the value of $a_S$ is now that
of $a_M$ and therefore is confined to be either $0$ or $1$.  This
means that we start with a flat potential in the weak coupling limit and
that as we move to the strong coupling limit a potential (perhaps
of sinusoidal shape)
develops and forces $a_S$ to take two possible values.  

From the physical point of view we would like the axion to be very small
and therefore, the value of $a_S=0$ should be picked by nature to the value
of $a_S=1$.   There is not apparent reason to distinguish between one or
the other in the model we have just analyzed, but we are tempted to 
say that perhaps the presence of smooth phase transitions and LGO
phases may play a role in destabilizing the second vacuum.
In any case, the existence of a potential implies that the axion picks up
a mass.  With our present knowledge of supermembranes, we are not able to
give an order of magnitude for the axion mass and therefore we are not
able to make any statement about the applicability of this axion 
potential to strong CP problem.

\section{The Winding Conjecture}

As explained in the introduction, we need a mechanism that will allow
us to treat multiple windings of membranes as multiple particle states
while treating multiple windings of the string as single particle states.
From the discussion of sections 3 and 4, we have learned that
the coupling of the string to the two-form potential is periodic and
the coupling of the membrane to the three-form potential is periodic 
and discrete.
The difference between these two couplings can be used to explain why
multiple windings of the membrane can be interpreted as multiple
particle states while multiple windings of the string can be
interpreted as single particle states.

\subsection{Model Dependent Degrees of Freedom in the Gauged Linear
Sigma Model}

A string propagating on a Calabi-Yau threefold does not have
gauge degrees of freedom which are model dependent. A membrane
can wrap about a two-cycle on the manifold and can thus have model
dependent gauge degrees of freedom.  For a compactification $T^2$,
both the membrane and the string can have model dependent gauge degrees
of freedom associated to the B-field and 
C-field respectively.  For the string they arise from wrapping the B-field
about the one cycles of $T^2$ while for the membrane they arises from
wrapping the C-field about the two cycle of $T^2$.  
The question is how these degrees of freedom arise in the string gauged linear
sigma model.  Clearly, the expression 
\be
\int d^2y \partial_i \phi^n\partial_j X^{\mu}\epsilon^{ij}B_{\mu n}
\ee
is not gauge invariant, and it is thus excluded from the
list of possible terms which can be added to the string gauged linear sigma
model.  However, we must remember that the string is 10 dimensional and
not two  dimensional.  Thus the  pull back of the  term
\be
\int d^2y \epsilon^{ij}\partial_j A_i=
\int d^2y \epsilon^{ij}B_{ij}\label{pb}
\ee
is not 
\be
\int d^2y \epsilon^{ij}\partial_j \phi^n \partial_i \phi^{m} B_{mn}
\label{PB}
\ee
which is the pull back to a conformally anomalous two (=dim($T^2$))
dimensional compact
target space with coordinates labeled by index $n,m$.  
Rather, for a target space $T^2\times R^8$ (\ref{pb}) takes the form
\be
\int d^2y \epsilon^{ij}(
\partial_j \phi^n \partial_i \phi^{m} B_{mn}+
\partial_j X^{\mu} \partial_i X^{\nu} B_{\mu\nu}+
\partial_j \phi^n \partial_i X^{\mu} B_{\mu n}),\label{f35}
\ee
which is the embedding of a string in a 10 dimensional space.
Thus, the model dependent gauge degrees of freedom associated to the
B-field
are found in the topological term of the string gauged linear sigma
model.

We can make a similar treatment of the topological term 
\be
\int d^3y \epsilon^{ijk}\partial_j A_{i} A_{k}=
\int d^3y \epsilon^{ijk}C_{ijk}\label{pb1}
\ee
found in the membrane  gauged linear sigma model.
The pull back of (\ref{pb1}) must include all the fields in the
11 dimensional space.  Thus, the pullback of (\ref{pb1}) for a
$T^2\times R^9$ target space takes the form
\be
\int d^3y \epsilon^{ijk}(
\partial_i \phi^m\partial_j \phi^n \partial_k X^{\mu} C_{mn\mu}+
\partial_i \phi^n\partial_j X^{\mu} \partial_k X^{\nu} C_{n \mu\nu}+
\partial_i X^{\mu} \partial_j X^{\nu} \partial_k X^{\rho} 
C_{\mu \nu\rho}).\label{rrr1}
\ee

Otherwise,  dimensional reduction of (\ref{rrr1}) would not
be consistent with (\ref{f35}) in the limit that 
$T^2\to S^1\times R$.

\subsection{Evidence for the Conjecture}

We continue 
by studying the different behaviors of the Dirac action for a membrane
wrapped once about an $S^1$ and a membrane wrapped $n$ times about the same
$S^1$.   The induced 
metric for the 11 dimensional background must be expressed
in terms of 10 dimensional backgrounds in order to carry out the
dimensional reduction \cite{D}.
This decomposition leads to an induced worldvolume metric of the form
\begin{equation}
\hat{g}_{\hat{i}\hat{j}}=
\Phi^{-2/3}\pmatrix{
g_{ij}+\Phi^2 A_iA_j & \Phi^2 A_i \cr
\Phi^2 A_j & \Phi^2 \cr
}.
\end{equation}
For this decomposition  and a membrane wrapped once about the $S^1$ we
find that $ \sqrt{\hat{g}}=\sqrt{g}$.  In the case that 
the membrane is wound $n$ times about $S^1$ we have instead
$ \sqrt{\hat{g}}=n \sqrt{g}$.  This is expected because $\sqrt{\hat{g}}$
measures the worldvolume, and by wrapping the membrane $n$ times about
$S^1$, the worldvolume is increased by a factor $n$.  We can
absorb this factor by simply redefining the worldvolume metric since
the membrane action is not Weyl invariant.

We could be tempted to interpret the $n$ in the worldvolume factor as
a rescaling of the membrane tension.  That is, {\it a priori} we
could use the argument that a membrane with $n$ units of  worldvolume and
tension $T$ has the same energy as a membrane  with a single 
unit of worldvolume
and tension $nT$ to argue that they lead to the same physics.
However, this statement is not true. The dimensional
reduction of these two membranes lead to strings with different physics.
The first case leads to a string with tension proportional to $T$ 
and $n$ units of  worldsheet.  The second case leads to a string with 
tension proportional to $nT$ and unit worldsheet.  In the first case we
can use the Weyl symmetry to get a string with unit worldsheet and tension
$T$.  This string has the same worldsheet as the string in the second
case but its 
tension is $n$ times larger than the first string.
Therefore, we cannot interpret the $n$ in the worldvolume
factor as a rescaling of the membrane tension.

From the paragraph above, we arrive to the conclusion that 
{\it the string tension
is not rescaled by the winding number of the membrane about the $S^1$ and
therefore a membrane wrapped $n$ times about $S^1$ does not lead to
a rescaling of the string tension, rather, it leads to a redefinition
of the worldvolume metric.}

We now take a look at the WZ term for the membrane.
The three form potential, $C$, appears in the membrane action in the term
\be
m\int d^3y \epsilon^{ijk}\partial_iX^{\mu}\partial_jX^{\nu}\partial_k\phi^{11}
 C_{11\mu\nu}\label{eq0}
\ee
where $m\in {\bf Z}$.  It is the pull back of the topological term
(\ref{nose}).
We consider the situation in which the membrane is wrapped $n$ times
about the one-cycle of $S^1$.  This is done by using the 
reparametrization 
\be
\phi^{11}=n y^3\label{eq1}
\ee
where $y^3$, a coordinate on the world volume,
 along with $\phi^{11}$ will be dimensionally reduced.  
Before doing any dimensional reduction we are already able to 
make a statement.   Substitution of (\ref{eq1}) into (\ref{eq0})
yields
\be
n m\int d^3y \epsilon^{ij3}\partial_iX^{\mu}\partial_jX^{\nu}
 C_{11\mu\nu}\label{eq2}
\ee
The coupling of the $C$ field after using reparametrization invariance
is rescaled
\be
m\to n m.
\ee
However,  $m$ is periodic.
Thus
\be
m\sim m+1
\ee
means that
\be
m n\sim m\label{p1}
\ee
and therefore, the axionic charge of the membrane is not renormalized
by the winding number of the membrane about $S^1$.

{\it This means that the membrane wrapping $n$ times about the $S^1$ does not
lead to a rescaling of the axionic charge. After dimensional reduction, 
the string  action obtained has unit axionic charge regardless of how 
many times the membrane wraps around the $S^1$.}

It is the integral nature of the coupling $m$ which has prevented the axionic
charge from being rescaled.
This is not the case for the string because the coupling of the
WZ term in the string action 
is not integral.  Then, the WZ term for the string
\be
\theta \int d^2y \epsilon^{ij}\partial_iX^{\mu}\partial_k\phi^{10}
 B_{10\mu}\label{eq3}
\ee
will have its coupling $\theta$ rescaled after the reparametrization
\be
\phi^{10}=n y^2\label{eq4}
\ee
Substitution of (\ref{eq4}) into (\ref{eq3}) gives
\be
n \theta \int dy \epsilon^{i2}\partial_iX^{\mu}
 B_{10\mu}\label{eq5}.
\ee
Therefore, the string coupling gets renormalized by the winding
number of the string about a one-cycle
\be
\theta\to n\theta.
\ee
For generic values of $\theta$ there is no identification
\be
\theta\sim n\theta
\ee
because $\theta$ is not an integer.
Thus the ``axionic" charge of space time particles depend on the number of 
times the string is wound about $S^1$, as expected.

These contrasting properties between $m$ and $\theta$ allow us to resolve the
problem of winding conjecture to a membrane wrapped about $T^2$.
Let us start with a membrane wrapped $n_1$ times about the first one-cycle of
$T^2$ and wrapped $n_2$ times about the second one-cycle of $T^2$.
We first perform a dimensional reduction of the WZ term for the membrane
about the first one-cycle.  This yields a WZ term for the string
\be
 m\int d^2y \epsilon^{ij3}\partial_iX^{\mu}\partial_jX^{\nu}
 C_{11\mu\nu}\label{eq6}
\ee
where we have used property (\ref{p1}).  We must now use the ansatz
\be
C_{11\mu\nu}=a(x^{\mu})B_{\mu\nu}
\ee
which takes the expression (\ref{eq6}) to the form
\be
\theta m\int d^2y \epsilon^{ij}\partial_iX^{\mu}\partial_jX^{\nu}
 B_{\mu\nu}\label{eq7}
\ee
where $\theta$ is the expectation value of $a(x^{\mu})$ which has
a periodic nature
\be
\theta\sim\theta+1
\ee
and therefore
\be
m \theta\sim \theta'\in [0,2\pi].
\ee
Thus, expression (\ref{eq6}) takes the final form
\be
\theta'\int d^2y \epsilon^{ij}\partial_iX^{\mu}\partial_jX^{\nu}
 B_{\mu\nu}\label{eq8}.
\ee
This expression is the WZ term for a string propagating in a
10 dimensional background.  Since the topology of the space is $R^9\times S^1$
we find that the WZ term is
\be
\theta'\int d^2y \epsilon^{ij}\partial_iX^{\mu}\partial_j\phi^{l}
 B_{\mu l}\label{eq9}.
\ee
As explained above, dimensional reduction of the string WZ term to a 
superparticle action  will depend on how many times the string in wound
about the $S^1$.
Thus, in carrying out the a second dimensional reduction about the
second one-cycle in order to obtain a particle, we find that the axionic charge
is renormalize by the winding of the string  and therefore  that
multiple winding of the string lead to different particle states,
in agreement with \cite{DH}.

\vspace{2cm}

{\bf Acknowledgements}

I am grateful to W. Israel, W. Lerche and 
E. Witten  for helpful discussion.
I would also
like thank McGill University, DAMTP, SISSA, ICTP, Ecole Polytechnique,
CEA (Saclay)  and CERN for kind hospitality during the summer.
This work is supported in part by NSERC Canada.

\pagebreak

\end{document}